\documentclass[prl,twocolumn,superscriptaddress]{revtex4-1}
\usepackage[dvipdfmx]{graphicx}
\usepackage{graphicx}
\usepackage{physics}
\usepackage{bm, amsmath, amssymb, braket}
\usepackage{times}
\usepackage{multirow}
\usepackage{ascmac}
\usepackage{amsthm}
\usepackage{float}
\usepackage{color}
\usepackage{comment}
\usepackage{mathtools}  
\usepackage{mathrsfs} 
\usepackage[colorlinks=True,urlcolor=blue,linkcolor=blue,citecolor=blue]{hyperref}
\usepackage{xcolor}
\usepackage{multirow}
\usepackage{setspace}
\usepackage{subfigure}
\usepackage[toc,page]{appendix}

\begin{document}
\title{Programmable Kondo Effect Formed by Landau Levels }
\author{Hong Chen}
\affiliation{National Laboratory of Solid State Microstructures and Department of Physics, Nanjing University, Nanjing 210093, China}
\author{Yun Chen}
\affiliation{National Laboratory of Solid State Microstructures and Department of Physics, Nanjing University, Nanjing 210093, China}
\author{Rui Wang}
\email{rwang89@nju.edu.cn}
\affiliation{National Laboratory of Solid State Microstructures and Department of Physics, Nanjing University, Nanjing 210093, China}
\affiliation{Collaborative Innovation Center of Advanced Microstructures, Nanjing University, Nanjing 210093, China}
\affiliation{Jiangsu Physical Science Research Center}
\affiliation{Hefei National Laboratory, Hefei 230088, People's Republic of China }
\author{Baigeng Wang}
\email{bgwang@nju.edu.cn}
\affiliation{National Laboratory of Solid State Microstructures and Department of Physics, Nanjing University, Nanjing 210093, China}
\affiliation{Collaborative Innovation Center of Advanced Microstructures, Nanjing University, Nanjing 210093, China}
\affiliation{Jiangsu Physical Science Research Center}

\begin{abstract}
Nanobubbles wield significant influence over the electronic properties of 2D materials, showing diverse applications ranging from flexible devices to strain sensors. Here, we reveal that a strongly-correlated phenomenon, \textit{i.e.}, Kondo resonance, naturally takes place as an intrinsic property of graphene nanobubbles. The localized strain within the nanobubbles engenders pseudo magnetic fields, driving pseudo Landau levels with degenerate Landau orbits.  Under the Coulomb repulsion, the Landau orbits form an effective $\mathrm{SU}(N)$ pseudospin coupled to the bath via exchange interaction. This results in a new flavor screening mechanism that drives an exotic flavor-frozen Kondo effect, which is absent in conventional Kondo systems.  The resonance here also exhibits an unparalleled tunability via strain engineering, establishing a versatile new platform to simulate novel correlated phenomena based on pseudo Landau levels.

\end{abstract}

\maketitle
\begin{figure}
\includegraphics[width=\linewidth]{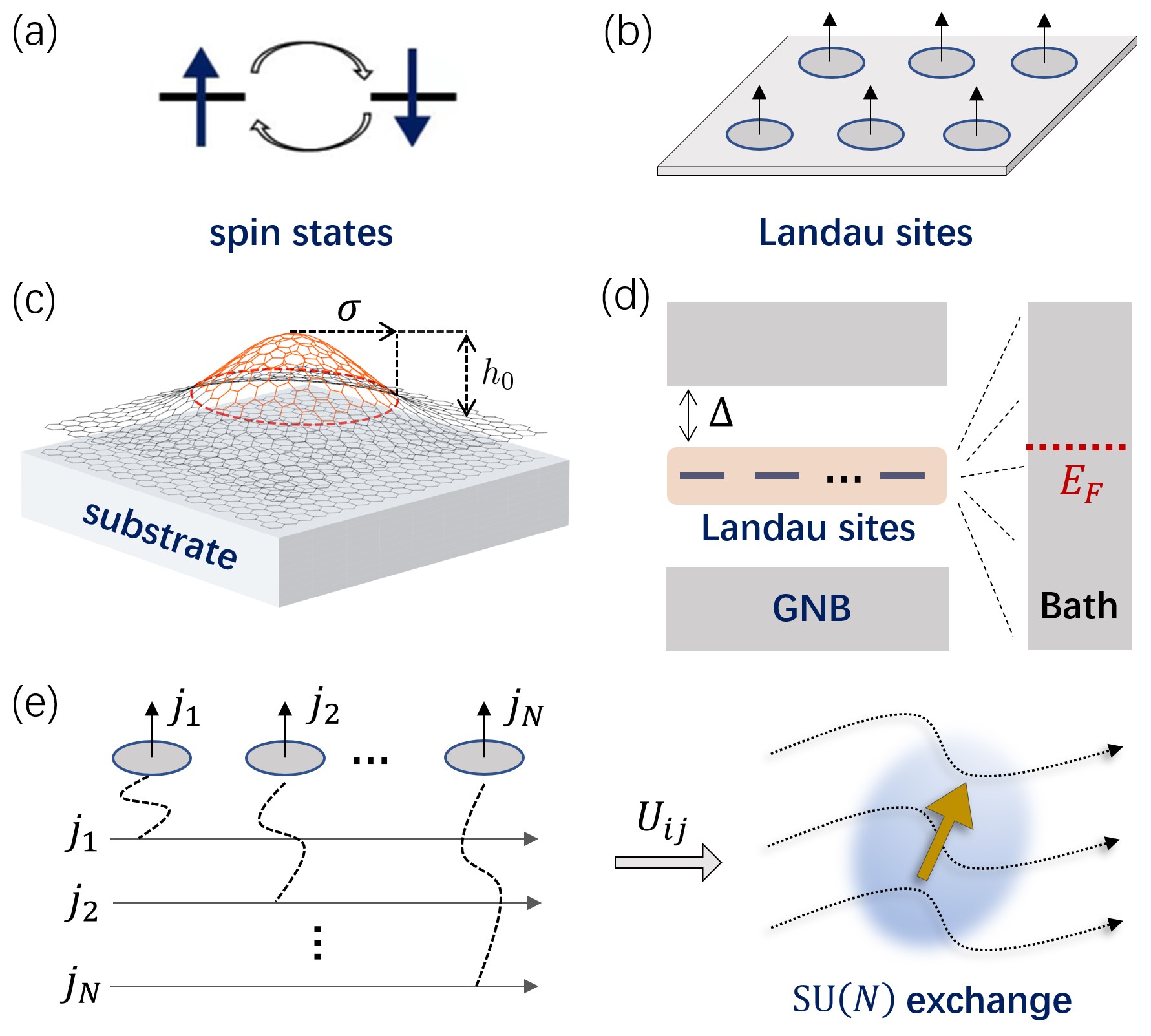}
\caption{\label{fig1} (a) The degenerate spin states constitute the key ingredient for Kondo effect. (b) Schematic plot of degenerate Landau orbitals in quantum Hall states, termed the LSs. (c) Plot of a bilayer graphene with a GNB on the top layer. (d) The LSs are separated by a gap from the higher energy GNB states, and are effectively coupled to the bath. (e) The flavor is conserved in the coupling, and an $\mathrm{SU}(N)$ pseudospin emerges from the LSs with considering the Coulomb repulsion.}
\end{figure}

\emph{\color{blue}{Introduction.--}} The Kondo effect stands as a cornerstone in condensed matter physics \cite{jkondo,Wilson}, sparking pivotal developments across diverse fields, ranging from impurity physics \cite{Balatsky} to quantum transport \cite{Meir,Grobis}. Its essence lies in the presence of local degenerate  quantum states (Fig.\ref{fig1}(a)) \cite{Beri}, typically realized through spin states in dilute magnetic alloys \cite{Hewson} or quantum dot junctions \cite{Cronenwett,Glazman}. The spin-exchange coupling with the host environment then plays a critical role in driving the many-body resonance, ultimately giving rise to the formation of Kondo singlet.

Manipulating the underlying symmetry of Kondo resonance is challenging \cite{Wenjie}, as it is dictated by intrinsic properties of the material, such as the degeneracy $N$ of the local quantum states. For example, extending the $\mathrm{SU}(2)$ Kondo model to $\mathrm{SU}(3)$ necessitates intentionally designed, intricate mesoscopic structures \cite{Carmi}, including carbon nanotubes \cite{Herrero,Makarovski,Tffang} and vertical quantum dots \cite{Sasaki}. Realizing the tunable $N$ $\mathrm{SU}(N)$ model is of crucial importance. Not only does this guarantee a variety of controllable Kondo phenomena \cite{Wwchen,Rzheng,Gchen,dma}, but also lays the foundation for realizing controllable Kondo lattices that host novel correlated physics \cite{Lacroix,Tsunetsugu,Tachiki,Alexandrov,Dzero}. 

In this Letter, we unveil exotic Kondo phenomena characterized by exceptional tunability and a fundamentally distinct mechanism. Our focus here is on Kondo behaviors emergent from nanobubbles \cite{zdai,Settnes,Pjia,Jlu,Kalashami} in two-dimensional (2D) materials \cite{Jren,Yzhang,linli,Hzhuang}, particularly graphene sheets. Graphene nanobubbles (GNB) represent smooth, dome-shaped deformations that induce localized strain \cite{Choi,Pereira1} (Fig.\ref{fig1}(c)). The strain generates pseudo magnetic fields within the bubble region \cite{Juan,Levy}, giving rise to a pseudo Landau level (LL) \cite{Castro,Rachel} with $N$ degenerate orbits, similar to that of the quantum Hall states (Fig.\ref{fig1}(b)). These orbits carry the ``flavor" quantum number, termed the Landau sites (LSs) \cite{Ezawa}. Here, we propose the LSs can be utilized as degenerate quantum states, the key ingredient for Kondo resonance, akin to the spin states (Fig.\ref{fig1}(a)). 

In realistic materials, the GNB is coupled to the bath, as illustrated by Fig.\ref{fig1}(c). This coupling manifests in low-energy as the hybridization between the LSs and bath electrons (Fig.\ref{fig1}(d)). Intriguingly, flavor conservation is maintained--each LS is hybridized with an effective 1D bath mode bearing the same flavor (Fig.\ref{fig1}(e)), resembling the Anderson impurity model \cite{Hewson}. Further considering the Coulomb interaction, an effective $\mathrm{SU}(N)$ pseudospin then emerges, generating the $\mathrm{SU}(N)$ Kondo resonance (Fig.\ref{fig1}(e)). Particularly, the interaction projected to Landau levels has a special form different from that of conventional Anderson models, resulting in an exotic flavor-frozen Kondo effect (FFKE) unique to our proposed platform. 

Recently, GNBs have been demonstrated to be programmable \cite{Pjia}. Particularly, the $N$-degeneracy can be manipulated by strain engineering. Thus, $\mathrm{SU}(N)$ Kondo models with successively tunable $N$ are realized in a single platform, establishing a flexible experimental framework for exploring Kondo physics  beyond conventional setups.

\textit{\color{blue}{Landau sites in GNBs.--}} We first analyze the strain effect in monolayers. The strain induces local lattice distortion. The three nearest-neighbor lattice vectors $\boldsymbol{\delta}_n$, with $n=1,2,3$, are modulated to $\mathbf{d}_n=\boldsymbol{\delta}_n+\hat{u}\boldsymbol{\delta}_n$, where $\hat{u}$ is the strain tensor whose matrix elements are $\hat{u}_{ij}=(\partial_i u_j+\partial_j u_i+\partial_iz\partial_jz)/2$, with $i,j=1,2$. Here, $\mathbf{u}=(u_1,u_2)$ and $z$ are respectively the in-plane and out-of-plane deformation. The three nearest-neighbor hoppings are accordingly modulated, \textit{i.e.}, $t_n=t_0e^{-\beta(|\mathbf{d}_n|/a-1)}$, where $t_0$ is the original hopping, and $\beta=3.37$ is used \cite{Carillo,Amal}. Consequently, the kinetic energy is modified at the $\mathbf{K}^{\pm}$ valley, as if the Dirac fermions were coupled to a pseudo gauge field $\mathbf{A}(\mathbf{r})$.  It satisfies $A_{x}(\mathbf{r})-iA_y(\mathbf{r})=\sum_n (t_n-t_0)e^{i\mathbf{K}^{\pm}\cdot\boldsymbol{\delta}_n}$ \cite{Pereira}, which not only applies to the in-plane but also to the out-of-plane strain \cite{Carillo}. The pseudo magnetic field, $B(\mathbf{r})=\nabla\times\mathbf{A}(\mathbf{r})$, is known to generate pseudo Landau levels \cite{Pereira, YinLJ}. For example, in the case of triaxial strain with $\mathbf{u}\propto(2xy,x^2-y^2)$, the corresponding $B(\mathbf{r})$ is uniform but opposite for the two valleys ($B$ and $-B$ for $\mathbf{K}^+$ and $\mathbf{K}^-$ respectively, and the flux $\Phi>0$ is assumed). LLs are then formed for both valleys with $E_n=\mathrm{sign}(n)v_F\sqrt{2|n|B}$ ($n=0,\pm1,...$).

A more natural description of GNBs is given by the out-of-plane Gaussian deformation \cite{Roldan_2015,Park_2023}, $z(\mathbf{r})=h_0e^{-r^2/2\sigma^2}$, where $\sigma$ and $h_0$ are respectively the radius and the height of the GNB (Fig.1(c)). In this case, the pseudo magnetic field is non-uniform \cite{sup}. Nevertheless, the averaged flux, $\overline{\Phi}=\int d\mathbf{r}B(\mathbf{r})$, becomes nonzero for $N=|N_B-N_A|\neq0$, where $N_A$ and $N_B$ are the number of A and B sublattices in the strained area. This leads to zero-energy LL with degeneracy $N$, as ensured by the Lieb-Sutherland theorem \cite{Lieb_1989,Sutherland,Inui}. Furthermore, the curvature of the bubble generates a weak spin-orbit coupling, whose effect can be safely neglected (Sec.8 of Supplemental Material).
\begin{figure}[!t]
    \centering    \includegraphics[width=\linewidth]{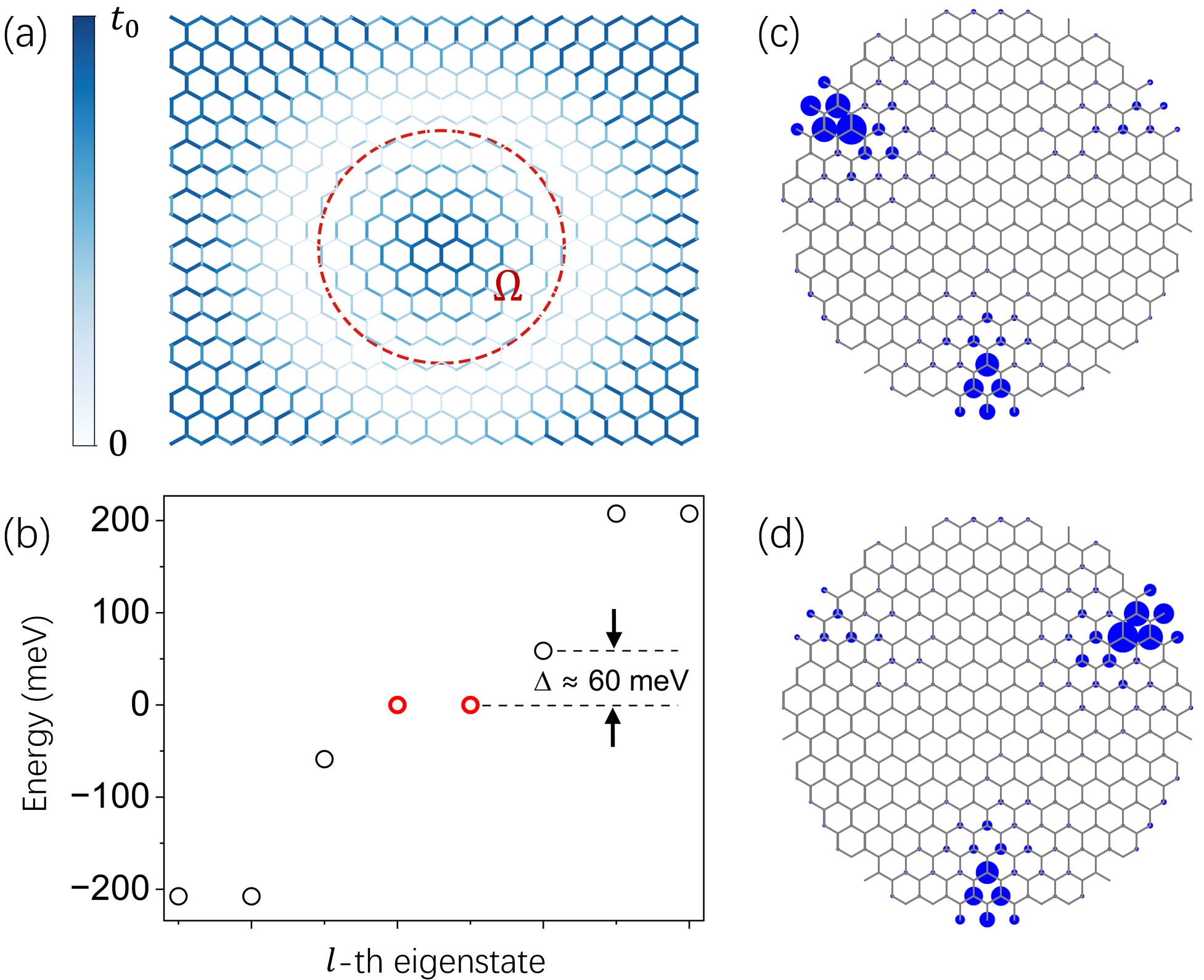}
    \caption{(a) The hopping $t_n$ for graphene with Gaussian deformation. The minima of $t_n$ lie on the red sphere, separating the strained GNB from the bath. $h_0=9.5a$, $\sigma=7a$, with $a$ being nearest-neighbor lattice constant. $N_A=61$, $N_B=63$ in $\Omega$.  (b) The low-energy spectrum of the GBN in (a).   Two zero modes are marked by red emerge, separated by a gap $\Delta$ from the higher energy states. (c),(d) The wavefunction distributions of the two zero modes.}
    \label{fig2}
\end{figure} 

For the Gaussian deformation, the atom separation in a unit cell first increases and then decreases with the distance from the strain center. This results in the hopping amplitude $t_n$ shown in Fig.\ref{fig2}(a), which decreases from $t_0$ to small values before increases up to $t_0$ at larger $r$. Thus, a strained subsystem ($\Omega$) is automatically separated and weakly coupled to the rest of the graphene sheet.  $\Omega$ is described by $H_{\mathrm{sys}}=-\sum_ {\mathbf{r}\in\Omega,n} t_n \psi^{\dagger}(\mathbf{r})\psi(\mathbf{r}+\boldsymbol{\delta}_n)+\mathrm{h.c.}$, where $\psi(\mathbf{r})$ is the annihilation operators of electrons at $\mathbf{r}$.  The corresponding energy spectrum is obtained and shown by Fig.\ref{fig2}(b). For $N=|N_B-N_A|$, we always find $N$ degenerate zero energy modes, in consistence with Lieb-Sutherland theorem. Notably, the zero modes are isolated from the higher energy states by a significant gap $\Delta$. For $N=2$ and the total site number $\sim100$, $\Delta\sim 60$meV$\simeq 696K$. Hence, the low-energy physics of the GNB is well captured by the zero modes.

For $N=2$, the wavefunction distributions of the two zero modes are shown in Fig.\ref{fig2}(c)-(d). Their analytic forms can be derived by solving the Liepmann-Schwinger equation describing the scattering of electrons by the non-uniform gauge potential $\mathbf{A}(\mathbf{r})$ \cite{Schneider}. In the symmetric gauge, the wavefunctions $\Psi^{(\xi)}_m(\mathbf{r})$ are sublattice polarized, and characterized by the orbital angular momentum (OAM) $m$ and the valley index $\xi=\pm$. For example, in the triaxial strain case with uniform $B$, $\Psi^{(\xi)}_m(\mathbf{r})$ is reduced to the typical Landau wavefunction, which, in the sublattice basis, reads as \cite{sup}, $\Psi^{(\xi)}_m(\mathbf{r})=[0,z_{\xi}^{m}(r/r_0)^{-N_{\Phi}}]^{\mathrm{T}}$ with $m=0,1,...,N_{\Phi}$, where $N_{\Phi}=\overline{\Phi}/2\pi$ is the number of flux quanta, $r_0$ is a constant, and $z_{\xi}=y+i\xi x$ is the complex coordinate. $\Psi^{(\xi)}_m(\mathbf{r})$ is fully polarized as its component is  nonzero only on B sublattice.

For convenience, we introduce the flavor, $j=(\xi,m)$ \cite{footnote0}, and represent the degenerate zero modes by $|j\rangle=d^{\dagger}_j|0\rangle$, where $|0\rangle$ denotes the vacuum, $d^{\dagger}_j$ is the creation operator satisfying, $\{d_{j},d^{\dagger}_{j^{\prime}}\}=\delta_{j,j^{\prime}}$. Despite the analogy with the LSs in quantum Hall physics \cite{Ezawa}, a key difference here is that the degeneracy is restricted by $N=|N_B-N_A|$.

\begin{figure}[t]
    \centering
    \includegraphics[width=\linewidth]{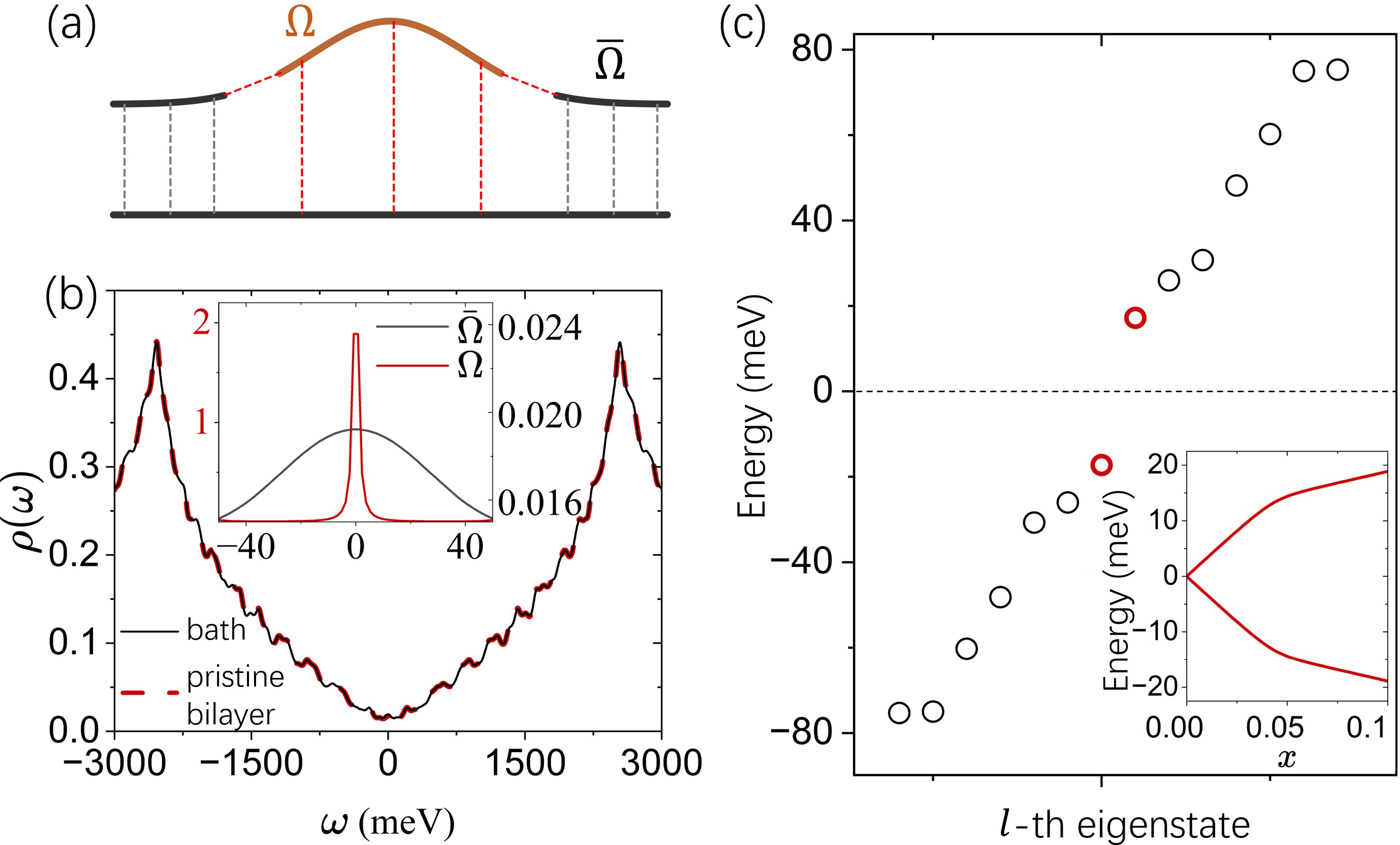}
    \caption{(a) The sideview of a bilayer graphene with a top layer GNB ($\Omega$) coupled to the bilayer environment $\overline{\Omega}$.  (b) The DOS of $\overline{\Omega}$ compared with that of a pristine bilayer, plot with a broadening factor of 2meV. The inset shows the separate DOS of $\Omega$ and $\overline{\Omega}$ around $\omega=0$, obtained from a tight-binding model with 800 unit cells for each layer.  (c)  The energy spectrum of the total system ($\Omega$+$\overline{\Omega}$) with their coupling, $t=xt_0$.  The shifted LSs are highlighted by red, and  $x=0.1$ is used. The inset shows the shift with increasing $x$. }
    \label{fig3}
\end{figure}

\textit{\color{blue}{Emergent $\mathrm{SU}(N)$ Anderson model.--}} The LSs in a GNB are inevitably subjected to interactions and hybridization with the hosting environment. To better investigate these effects, we consider Bernal-stacked bilayer graphene with a GNB (Fig.\ref{fig3}(a)). Starting from the tight-binding model describing the GNB ($\Omega$) coupled to the bilayer environment ($\overline{\Omega}$), we will derive a low-energy effective theory of the LSs in hybrization with the bath electrons.

We first calculate the density of states (DOS) of $\Omega$. As shown by the red curve in the inset to Fig.\ref{fig3}(b), a zero-energy DOS peak is found. This reflects the contribution of the zero-energy LSs, formally described by $H_{\mathrm{d}}=\sum^N_{j=1}\epsilon_dd^{\dagger}_jd_j$, where $\epsilon_d=0$.  Then, we simulate $\overline{\Omega}$ by a graphene bilayer with the GNB being cut out. As shown by Fig.\ref{fig3}(b), its DOS appears almost the same as that of the perfect bilayer graphene. The similarity maintains in low-energy, only exhibiting a slightly enhanced DOS around $\omega=0$ compared to the perfect bilayer (Sec.2 of the Supplemental Material). Thus, as shown by the inset to Fig.\ref{fig3}(b), we arrive at a low-energy bath continuum hosting the discrete zero-energy LSs.

To visualize the hybridization between the LSs and bath continuum, we diagonalize the total system. Comparing the energy spectrum (Fig.\ref{fig3}(c)) and that of the GNB alone (Fig.\ref{fig2}(b)), we find that the LSs are shifted away from $E=0$. The shift becomes more  significant with increasing the system-environment coupling (inset to Fig.\ref{fig3}(c)). This reveals the hybridization between the LSs and the bath (Fig.\ref{fig1}(d)), a prerequisite for Kondo resonance. Since the Kondo resonance is quantitatively determined by the bath DOS around the Fermi energy \cite{Hewson}, replacing the bath by the perfect graphene bilayer can facilitate our calculations without affecting the Kondo behaviors.

The hybridization is measured by the overlap, $V=\langle j|\phi^{(\xi)}_{\mathbf{k},A}\rangle$, where $|\phi^{(\xi)}_{\mathbf{k},A}\rangle$ is the Bloch wavefunction on A sublattice of the bilayer \cite{footnote00}. In the band basis, $|\phi^{(\xi)}_{\mathbf{k},A}\rangle$ is represented by, $|\phi^{(\xi)}_{\mathbf{k},A}\rangle=(-e^{-i2\xi\theta}|c^{(\xi)}_{\mathbf{k},+}\rangle+e^{i2\xi\theta}|c^{(\xi)}_{\mathbf{k},-}\rangle)/\sqrt{2}$, where $|c^{(\xi)}_{\mathbf{k},\pm}\rangle$ denotes the Bloch state of conduction/valance band electrons at valley $\xi$, and $\theta$ is the angle of $\mathbf{k}$. Expanding $|c^{(\xi)}_{\mathbf{k},\pm}\rangle$ by the OAMs, \textit{i.e.}, $u^{(\xi)}_{\mathbf{k},\pm}\equiv\langle\mathbf{r}|c^{(\xi)}_{\mathbf{k},\pm}\rangle=\sum_{m}u^{(\xi)}_{k,m,\pm}e^{im\theta}$, the hybridization between the LSs and the conduction/valence electrons is then obtained as, $V_{\pm}=\mp\delta_{\xi \xi^{\prime}}\delta_{m^{\prime},m\xi-2\xi}v_{k,\xi,m^{\prime}}/\sqrt{2}$, where $v_{k,\xi,m}=\int^{\infty}_0 drr^m(r/r_0)^{-N_{\Phi}}u^{(\xi)}_{k,m,\pm}(r)$.

Interestingly, two Kronecker-delta functions occur in $V_{\pm}$, enforcing the valley and the OAM conservation. In addition, due to the sublattice-momentum locking \cite{McCann}, the bath OAM, $m$, is shifted by 2 in the hybridization \cite{sup}. In the second-quantized form, the above hybridization process is described by,
\begin{equation}\label{eqhyb}
H_{\mathrm{hyb}}=\int dk\sum_{j} v_{k,j}d^{\dagger}_{j}(-c_{k,j,+}+c_{k,j,-}),
\end{equation}
where the $c$-fermions denote the bath electrons with shifted OAMs. Similarly, the bath Hamiltonian is reduced to,
 \begin{equation}\label{eqbath}
H_{\mathrm{bath}}=\int dk\sum_{j,n} (\epsilon_{k,n}-\mu) c^{\dagger}_{k,j,n}c_{k,j,n},
 \end{equation}
where $\epsilon_{k,n}=nk^2/2m$, with $n=\pm$. We have introduced the chemical potential $\mu$ of the bath, which can be tuned by the gate voltage.

Further introducing effective fermionic operators in the energy representation that combine the conduction and valence bands \cite{rwang,zitko}, \textit{i.e.}, $\gamma_{\epsilon,j}=[-c_{\epsilon,j,+}\theta(\epsilon)+c_{\epsilon,j,-}\theta(-\epsilon)]/2$ and $\gamma^{\dagger}_{\epsilon,j}=[-c^{\dagger}_{\epsilon,j,+}\theta(\epsilon)+c^{\dagger}_{\epsilon,j,-}\theta(-\epsilon)]/2$, we finally arrive at,
\begin{equation}\label{eqbath2}
  H_{\mathrm{bath}}=\sum^N_{j=1} \int^D_{-D} d\epsilon(\epsilon-\mu) \gamma^{\dagger}_{\epsilon,j}\gamma_{\epsilon,j},
\end{equation}
and
\begin{equation}\label{eqhyb2}
H_{\mathrm{hyb}}=\sum^N_{j=1}\int^D_{-D} d\epsilon [h_j(\epsilon)d^{\dagger}_j\gamma_{\epsilon,j}+h.c.],
\end{equation}
where $D$ is the energy cutoff, $h_j(\epsilon)=|v_{\epsilon,j}|^2\rho_{\overline{\Omega}}(\epsilon)$ is the broadening function with $\rho_{\overline{\Omega}}(\epsilon)$ being the bath DOS. Eq.\eqref{eqhyb2} implies that the zero-energy LSs are only coupled to the bath states with the same flavor $j$. Furthermore, Eq.\eqref{eqbath} indicates that the bath states of flavor $j$ are essentially fermionic modes exhibiting 1D momentum $k$. Thus, we reveal a hidden $N$-flavor 1D hybridization model underlying the GNB, as demonstrated by Fig.\ref{fig1}(e).

Furthermore, the electron-electron interaction is non-negligible in the GNB \cite{footnote1}, $H_{\mathrm{int}}=\int d\mathbf{r}d\mathbf{r}^{\prime}U(\mathbf{r}-\mathbf{r}^{\prime})\psi^{\dagger}(\mathbf{r})\psi(\mathbf{r})\psi^{\dagger}(\mathbf{r}^{\prime})\psi(\mathbf{r}^{\prime})$, where $U(\mathbf{r}-\mathbf{r}^{\prime})$ is short-ranged due to the strong screening in the bilayer \cite{xfwang}. Projecting $H_{\mathrm{int}}$ to the zeroth LL leads to \cite{sup},
\begin{equation}\label{eqint}
  H_{\mathrm{int}}=\sum_{i\neq j} U_{ij}d^{\dagger}_{i}d_id^{\dagger}_jd_j,
\end{equation}
where $U_{ij}$ is the direct integral between the LSs with the flavors, $i=(\xi^{\prime},m^{\prime})$ and $j=(\xi,m)$. For the hardcore repulsion, $U_{ij}$ is evaluated as \cite{Ezawa}, $U_{ij}=(m+m^{\prime})!/8\pi l_B2^{m+m^{\prime}}m!m^{\prime}!$, where $l_B$ is the magnetic length. Remarkably, Eq.\eqref{eqbath2}-Eq.\eqref{eqint} constitute an emergent $\mathrm{SU}(N)$ Anderson impurity model that hosts Kondo effect, as shown by the numerical renormalization group (NRG) calculations (Sec.3.3,Sec.4 of Supplemental Materials \cite{sup}), which is double verified by exact diagonalization of finite-size lattice models (Sec.5 of Supplemental Materials \cite{sup}).

\begin{figure}[t]
    \centering
    \includegraphics[width=\linewidth]{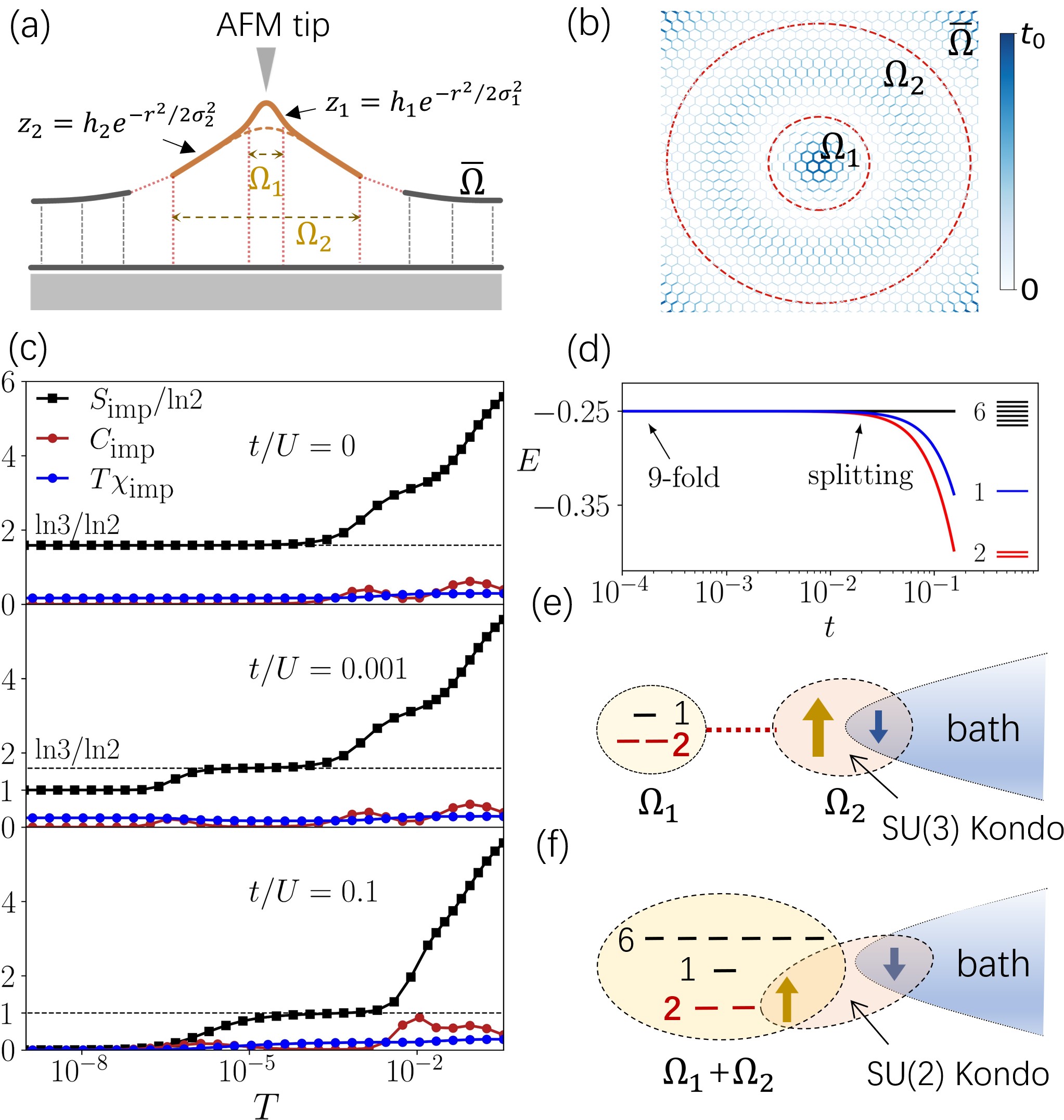}
    \caption{ (a) The sideview of a graphene bilayer with concentric Gaussian deformation, which realizes a side-coupled two-impurity Anderson model. (b) The calculated hopping amplitudes on the upper layer. (c) The temperature-dependent thermodynamics of the side-coupled Anderson model with $U=1/8\pi l_B$. (d) The exact diagonalization spectrum of the impurity system. (e) accounts for the residue ln2-entropy for intermediate $t$, and (f) illustrates the physical mechanism of the FFKE. }
    \label{fig4}
\end{figure}

\textit{\color{blue}{Flavor-frozen Kondo effect.}}--We now explore a more exotic Kondo phenomenon unique to the GNBs. We consider two concentric Gaussian deformations, $z_1(\mathbf{r})=h_1e^{-r^2/2\sigma^2_1}$ and $z_2(\mathbf{r})=h_2e^{-r^2/2\sigma^2_2}$ with $\sigma_1<\sigma_2$, which can be realized by modulating the local strain on top of a nanobubble using the functional atomic force microscopy (AFM) tip  (Fig.\ref{fig4}(a)).  Such a deformation induces two concentric strained regions, $\Omega_n$ with $n=1,2$, each supporting $N_n=|N_{n,B}-N_{n,A}|$ LSs.  As indicated by Fig.\ref{fig4}(b), $\Omega_1$ is weakly coupled to $\Omega_2$, which is in turn coupled to the bath $\overline{\Omega}$. 

Following the same approach, we derive a side-coupled two-impurity Anderson model (Sec.6 of supplemental material \cite{sup}), $H_{\mathrm{tot}}=H_{\mathrm{imp},1}+H_{\mathrm{imp},2}+H_{\mathrm{c}}+H_{\mathrm{hyb}}+H_{\mathrm{bath}}$, where $H_{\mathrm{imp},n}=\sum_{j\neq j^{\prime}}U_{ij}d^{\dagger}_{n,i}d_{n,i}d^{\dagger}_{n,j}d_{n,j}$ is the projected interaction within $\Omega_n$ ($n=1,2$). As will be clear below, the unique form of $U_{ij}$ distinguishes the Landau-level based Kondo physics from conventional ones in magnetic impurities or nanostructures.  $H_{\mathrm{c}}=t\sum^{N_j}_{j=1}(d^{\dagger}_{1,j}d_{2,j}+h.c.)$ is the coupling between LSs in $\Omega_1$ and $\Omega_2$, and $N_j=\mathrm{min}(N_1,N_2)$. The bath, $H_{\mathrm{bath}}$, and its hybridization with the LSs in $\Omega_2$, $H_{\mathrm{hyb}}$, are of the same form as Eq.\eqref{eqbath2} and Eq.\eqref{eqhyb2}.

The numbers of LSs, $N_1$ and $N_2$, are tunable by strain engineering. We first consider the $N_1=N_2=3$ case, and calculate the temperature-dependent thermodynamics using NRG \cite{Bulla} (Fig.\ref{fig4}(c)). For $t=0$, the ``impurity" entropy, i.e., the entropy of all the LSs in $\Omega_1$ and $\Omega_2$, flows to $\mathrm{ln}3$  at low temperatures. This indicates that the emergent SU(3) pseudospin of $\Omega_2$ is fully screened by the bath, while the three-fold degenerate LSs in $\Omega_1$ remains intact. For intermediate $t$ (middle panel of Fig.\ref{fig4}(c)), despite the $\mathrm{ln}3$ plateau, the system further flows to an unexpected fixed point with $\mathrm{ln}2$-entropy.  Further enlarging $t$ (lower panel of Fig.\ref{fig4}(c)), the $\mathrm{ln}3$-plateau vanishes, leaving a $\mathrm{ln}2$-plateau, which finally flows to zero entropy at low temperatures. Meanwhile, the local moment $T\chi$ also flows to zero, indicating that there are no free flavors of LSs left.

To understand the thermodynamics, we turn off the coupling between $\Omega_2$ and $\overline{\Omega}$, and diagonalize the impurity system, i.e., $H_{\mathrm{imp},1}+H_{\mathrm{imp},2}+H_{\mathrm{c}}$. The lowest energy spectrum as a function of $t$ is plot in Fig.\ref{fig4}(d). With gradually increasing $t$, the originally 9-fold degenerate ground state (3-fold degeneracy for both $\Omega_1$ and $\Omega_2$) split into three branches of spectrum, with 6-, 1-, 2-fold degeneracy, respectively. Such a 6-1-2 splitting behavior originates from the projected interaction $U_{ij}$, and is absent in conventional coupled Anderson models. 

Fig.\ref{fig4}(d) reveals an interesting FFKE unique to the GNB. For intermediate $t$, although the SU(3) Kondo screening takes place between $\Omega_2$ and bath, the Kondo cloud still has couplings with the LSs in $\Omega_1$, leading to its splitting. Thus, at low-temperatures, the original 3 flavors in $\Omega_1$ are frozen down to 2, as marked by the red levels in Fig.\ref{fig4}(e). This accounts for the residual ln2-entropy. For large $t$, as indicated by Fig.\ref{fig4}(f), $\Omega_1$ and $\Omega_2$ are strongly hybridized, exhibiting  the fully split spectrum in Fig.\ref{fig4}(d), before the coupling to the bath sets in. Thus, the 2-fold degenerate ground states act as an effective SU(2) pseudospin, further screened by the bath at low-temperatures. Therefore, a remarkable SU(2) Kondo singlet ground state emerges from the side-coupled SU(3) Anderson model, due to the flavor-freezing of LSs.

\textit{\color{blue}{Conclusion and discussion.}}-- The Kondo behavior found here is formed by LSs, in contrast with the Kondo effect arising from the vacancy-induced local moments in bipartite lattices \cite{Lieb_1989,jhchen}. The key difference here is that the true spins are not involved in the screening process, similar to the charge Kondo effect \cite{Iftikhar,Matsushita, tfangn} realized in nanostructures \cite{Jesper}. The Kondo cloud here, essentially a ``flavor singlet", would be immune to measurements that probe spin or magnetic properties, such as the nuclear magnetic resonance spectroscopy. However, the transport anomalies, such as the resistivity minima, maintain to characterize the collective Fermi liquid behavior \cite{Mora} at the strong-coupling fixed point. This mechanism is not limited to graphene, but also applicable to other 2D materials exhibiting strain-induced quantization \cite{Zhao_2020,Gastaldo_2023,Qi_2023}. 

GNBs have been recently demonstrated to be programmable \cite{Pjia}. Using AFM, the size, shape, location and the number of GNBs can all be controlled. This offers an unparalleled platform to realize Landau level-based Kondo physics and their ``evolutions" with tuning $N$.  For example, in the concentric Gaussian deformation case above, we find that the FFKE evolves to a two-stage Kondo effect, with tuning $N_1=N_2=3$ to $N_1=N_2=2$  (Sec.7 of Supplemental Material). More interestingly, since spins are not involved in the flavor screening, they could be utilized as additional channels, if sufficient Rashba spin-orbit coupling and Zeeman fields are further introduced (Sec.9 of Supplemental Material). This offers new opportunities to explore exotic non-Fermi liquid physics, via simulating multi-channel Kondo models based on pseudo Landau levels.



\begin{acknowledgments}
H. C and Y. C. contributed equally to this work.

R. W. acknowledges Tigran Sedrakyan, Shaochun Li, Jianpeng Liu, Wei Su  for fruitful discussions. This work was supported by the Innovation Program for Quantum Science and Technology (Grant no. 2021ZD0302800), the National R\&D Program of China (2022YFA1403601), the National Natural Science Foundation of China (No. 12322402, No. 12274206), the Natural Science Foundation of Jiangsu Province (No. BK20233001), and the Xiaomi foundation.
\end{acknowledgments}


\begin{thebibliography}{99}
\bibitem{jkondo} J. Kondo, Prog. Theor. Phys. \textbf{32}, 37 (1964).

\bibitem{Wilson} K. G. Wilson, Rev. Mod. Phys. \textbf{47}, 773 (1975).

\bibitem{Balatsky} A. V. Balatsky, I. Vekhter, and J.-X. Zhu, Rev. Mod. Phys. \textbf{78}, 373 (2006).

\bibitem{Meir} Y. Meir, K. Hirose, and N. S. Wingreen, Phys. Rev. Lett. \textbf{89}, 196802 (2002).

\bibitem{Grobis} M. Grobis, I. G. Rau, R. M. Potok, H. Shtrikman, and D. Goldhaber-Gordon,  Phys. Rev. Lett. \textbf{100}, 246601 (2008).

\bibitem{Beri} B. Béri and N. R. Cooper,  Phys. Rev. Lett. \textbf{109}, 156803 (2012).

 \bibitem{Hewson} A. C. Hewson, \textit{The Kondo Problem to Heavy Fermions}, (Cambridge University Press, Cambridge, England, 1993).

 \bibitem{Cronenwett} S. M. Cronenwett, T. H. Oosterkamp, and L. P. Kouwenhoven, Science \textbf{281}, 540 (1998).
 
\bibitem{Glazman} M. Pustilnik and L. Glazman,  J. Phys.: Condens. Matter \textbf{16}, R513  (2004).

\bibitem{Wenjie} W. Liang, M. P. Shores, M. Bockrath, J. R. Long, and H. Park, Nature \textbf{417}, 725 (2002) . 

\bibitem{Carmi} A. Carmi, Y. Oreg, and M. Berkooz, Phys. Rev. Lett. \textbf{106}, 106401 (2011).

\bibitem{Herrero} P. Jarillo-Herrero, J. Kong, H. S. J. van der Zant, C. Dekker, L. P. Kouwenhoven, and S. De Franceschi, Nature \textbf{434}, 484
(2005).

\bibitem{Makarovski}  A. Makarovski, A. Zhukov, J. Liu, and G. Finkelstein, Phys. Rev. B \textbf{75}, 241407(R) (2007).

\bibitem{Tffang} T.-F. Fang, W. Zuo, and H.-G. Luo, Phys. Rev. Lett. \textbf{101}, 246805 (2008). 

\bibitem{Sasaki} S. Sasaki, S. Amaha, N. Asakawa, M. Eto, and S. Tarucha, Phys. Rev. Lett. \textbf{93}, 017205 (2004).

\bibitem{Wwchen} W. Chen, Y. Yan, M. Ren, T. Zhang, and D. Feng, Sci. China Phys. Mech. Astron. \textbf{65}, 246811 (2022).  

\bibitem{Rzheng} R. Zheng, R.-Q. He and Z.-Y. Lu, Chinese Phys. Lett. \textbf{35}, 067301 (2018). 

\bibitem{Gchen} G. Chen and J. L. Lado, Phys. Rev. Research \textbf{2}, 033466 (2020).

 \bibitem{dma} D. Ma, H. Chen, H. Liu, and X. C. Xie, Phys. Rev. B \textbf{97}, 045148 (2018).

\bibitem{Lacroix} C. Lacroix and M. Cyrot, Phys. Rev. B \textbf{20}, 1969 (1979).

\bibitem{Tsunetsugu} H. Tsunetsugu, M. Sigrist, and K. Ueda, Rev. Mod. Phys. \textbf{69}, 809 (1997).

\bibitem{Tachiki} M. Tachiki and S. Maekawa, Phys. Rev. B \textbf{29}, 2497 (1984).

\bibitem{Alexandrov} V. Alexandrov, P. Coleman, and O. Erten, Phys. Rev. Lett. \textbf{114}, 177202 (2015).

\bibitem{Dzero} M. Dzero, K. Sun, V. Galitski, and P. Coleman, Phys. Rev. Lett. \textbf{104}, 106408 (2010).

\bibitem{zdai} Z. Dai, Y. Hou, D. A. Sanchez, G. Wang, C. J. Brennan, Z. Zhang, L. Liu, and N. Lu, Phys. Rev. Lett. \textbf{121}, 266101 (2018).

\bibitem{Settnes} M. Settnes, S. R. Power, M. Brandbyge, and A.-P. Jauho, Phys. Rev. Lett. \textbf{117}, 276801 (2016).

\bibitem{Pjia} P. Jia, W. Chen, J. Qiao, M. Zhang, X. Zheng, Z. Xue, R. Liang, C. Tian, L. He, Z. Di, and X. Wang, Nat. Commun. \textbf{10}, 3127 (2019).

\bibitem{Jlu} J. Lu, A.H. C. Neto, and K. P. Loh,  Nat. Commun. \textbf{3}, 823 (2012) 

\bibitem{Kalashami} H. G.-Kalashami, K. S. Vasu, R. R. Nair, F. M. Peeters, and M. N.-Amal, Nat. Commun. \textbf{8}, 15844 (2017).

\bibitem{Jren}  J. Ren, H. Guo, J. Pan, Y. Yang, Z, X. Wu, H.-G. Luo, S. Du, S. T. Pantelides, and H.-J. Gao, Nano Lett. \textbf{14},  4011 (2014).
 
\bibitem{Yzhang} Y. Zhang, L. Li, J.-Hua Sun, D.-Hui Xu, R. L\"{u}, H.-G. Luo, and W.-Q. Chen, 
Phys. Rev. B \textbf{101}, 035124 (2020),

 \bibitem{linli} L. Li, Y.-Y. Ni, Y. Zhong, T.-F. Fang and H.-G. Luo,  New J. Phys. \textbf{15}, 053018 (2013).

 \bibitem{Hzhuang} H.-B. Zhuang, Q.-F. Sun and X. C. Xie, Euro. Phys. Lett. \textbf{86}, 58004 (2009). 


\bibitem{Choi} S.-M. Choi, S.-H. Jhi, and Y.-W. Son, Phys. Rev. B \textbf{81}, 081407(R) (2010),

\bibitem{Pereira1} V. M. Pereira, A. H. Castro Neto, and N. M. R. Peres, Phys. Rev. B \textbf{80}, 045401 (2009).

\bibitem{Juan} F. de Juan, J. L. Ma\~{n}es, and M. A. H. Vozmediano, Phys. Rev. B \textbf{87}, 165131 (2013).

\bibitem{Levy} N. Levy, S. A. Burke, K. L. Meaker, M. Panlasigui, A. Zettl, F. Guinea, A. H. Castro Neto, and M. F. Crommie, Science \textbf{329}, 544 (2010).

\bibitem{Castro} E. V. Castro, M. A. Cazalilla, and M. A. H. Vozmediano, Phys. Rev. B \textbf{96}, 241405(R) (2017).

\bibitem{Rachel} S. Rachel, I. G\"{o}thel, D. P. Arovas, and M. Vojta, Phys. Rev. Lett. \textbf{117}, 266801 (2016).

\bibitem{Ezawa} Z. F. Ezawa, \textit{Quantum Hall effects: Field theoretical approach and related topics} (World Scientific Publishing Company, 2008).

\bibitem{Carillo} R. C. Bastos, M. Ochoa, S.A. Zavala, and F. Mireles, Phys. Rev. B \textbf{98}, 165436 (2018).

\bibitem{Amal} M. N. Amal, L. Covaci, K. Shakouri, and F. M. Peeters, Phys. Rev. B \textbf{88}, 115428 (2013).

\bibitem{Pereira}  V. M. Pereira, and , A. H. Castro Neto, Phys. Rev. Lett. \textbf{103}, 046801 (2009).

\bibitem{YinLJ}  L.-J. Yin, S.-Y. Li,  J.-B. Qiao, J.-C. Nie, and L. He, Phys. Rev. B \textbf{91}, 115405 (2015)

\bibitem{Roldan_2015} R. Rold\'{a}n, A. Castellanos-Gomez, E. Cappelluti, and F. Guinea, J. Phys. Condens. Matter \textbf{27}, 313201 (2015).

\bibitem{Park_2023} H. C. Park, J. Y. Han, and N. Myoung, Quantum Sci. Technol. \textbf{8}, 025012 (2023).

\bibitem{sup} See supplemental materials for pertinent technical details on relevant proofs and derivations, which include Ref.\cite{Rozhkov,Ezawa,rwang,zitko,Bulla,Daniel}.

\bibitem{Rozhkov} A. V. Rozhkov, A. O. Sboychakov, A. L. Rakhmanov, and F. Nori, Phys. Rep. \textbf{648}, 1 (2016).

\bibitem{Daniel} Daniel Huertas-Hernando, F. Guinea, and Arne Brataas, Phys. Rev. B \textbf{74}, 155426 (2006).

\bibitem{Lieb_1989} E. H. Lieb, Phys. Rev. Lett. \textbf{62}, 1201 (1989).

\bibitem{Sutherland} B. Sutherland, Phys. Rev. B \textbf{34}, 5208 (1986).

\bibitem{Inui} M. Inui, S. A. Trugman, and E. Abrahams, Phys. Rev. B \textbf{49},
3190 (1994).

\bibitem{Schneider} M. Schneider, D. Faria, S. Viola Kusminskiy, and N. Sandle, Phys. Rev. B \textbf{91}, 161407(R) (2015)

\bibitem{footnote0} Note that $j$ should be understood as a general quantum number labeling the LS degeneracy, and it does not refer to the OAM for small bubbles with irregular shapes. 

\bibitem{footnote00}  Note that $\langle j|\phi^{(\xi)}_{\mathbf{k},B}\rangle$ does not have contribution because $|j\rangle$ (fully-polarized on B) is connected to A sublattices via  nearest neighbor hoppings.


\bibitem{McCann} E. McCann and M. Koshino, Rep. Prog. Phys. \textbf{76}, 056503 (2013).

\bibitem{rwang} R. Wang, W. Su, J.-X. Zhu, C. S. Ting, H. Li, C. Chen,
B. Wang, and X. Wang, Phys. Rev. Lett. \textbf{122}, 087001 (2019)

\bibitem{zitko} R. \v{Z}itko, Phys. Rev. B \textbf{81}, 241414(R) (2010).

\bibitem{footnote1} The Coulomb interaction becomes quite dominant in the GNB due to the presence of degenerate LSs. However, it can be omitted in the bath, because it merely leads to Fermi liquid corrections. Note that unlike the pristine bilayer graphene in thermodynamic limit, the bath  $\overline{\Omega}$ does not possess quadratic band touching points at $E=0$. Thus, the Coulomb interaction is irrelevant in  the bath. 

\bibitem{xfwang}  X.-F. Wang and T. Chakraborty, Phys. Rev. B \textbf{75}, 041404(R) (2007).

\bibitem{Bulla} R. Bulla, T. A. Costi, and T. Pruschke, Rev. Mod. Phys. \textbf{80},
395 (2008).

\bibitem{jhchen} J.-H. Chen, L. Li, W. G. Cullen, E. D. Williams and M. S. Fuhrer, Nat. Phys. \textbf{7}, 535 (2011).

\bibitem{Iftikhar} Z. Iftikhar, S. Jezouin, A. Anthore, U. Gennser, F. D. Parmentier, A. Cavanna, and F. Pierre, Nature \textbf{526}, 233 (2015).

\bibitem{Matsushita} Y. Matsushita, H. Bluhm, T. H. Geballe, and I. R. Fisher, Phys. Rev. Lett. \textbf{94}, 157002 (2005).

\bibitem{tfangn} T.-F. Fang, A.-M. Guo, H.-T. Lu, H.-G. Luo, and Q.-F. Sun, Phys. Rev. B \textbf{96}, 085131.

\bibitem{Jesper} Jesper Nygård, David Henry Cobden, and Poul Erik Lindelof,  Nature \textbf{408}, 342 (2000).

\bibitem{Mora} C. Mora, Phys. Rev. B \textbf{80}, 125304 (2009).

\bibitem{Gastaldo_2023} M. Gastaldo, J. Varillas, Á. Rodríguez, M. Velický, O. Frank, and M. Kalbáč, npj 2D Mater. Appl. \textbf{7}, 71 (2023).

\bibitem{Qi_2023} Y. Qi, M. A. Sadi, D. Hu, M. Zheng, Z. Wu, Y. Jiang, and Y. P. Chen, Adv. Mater. \textbf{35}, 2205714 (2023).

\bibitem{Zhao_2020} C. Zhao, M. Hu, J. Qin, B. Xia, C. Liu, S. Wang, \textit{et al.}, Phys. Rev. Lett. \textbf{125}, 046801 (2020).
















\end{thebibliography}
\end{document}